\documentclass[12pt]{article}
\usepackage{amsmath,amsfonts, feynmp, epsf}
\usepackage{amssymb}
\usepackage{cancel}
\usepackage{graphicx}
\usepackage{grffile}
\usepackage{array}
\usepackage{makecell}
\newcolumntype{x}[1]{>{\centering\arraybackslash}p{#1}}
\input epsf

\def\be{\begin{equation}}
\def\ee{\end{equation}}
\def\bea{\begin{eqnarray}}
\def\eea{\end{eqnarray}}
\def\ie{\begin{equation}\begin{aligned}}
\def\fe{\end{aligned}\end{equation}}

\newcommand{\A}{{\alpha}}
\newcommand{\B}{{\beta}}
\newcommand{\C}{{\gamma}}
\newcommand{\D}{{\delta}}

\makeatletter\@addtoreset{equation}{section}\makeatother

\newcommand{\cO}{{\mathcal O}}

\renewcommand{\title}[1]{\vbox{\center\LARGE{#1}}\vspace{5mm}}
\renewcommand{\author}[1]{\vbox{\center#1}\vspace{5mm}}
\newcommand{\address}[1]{\vbox{\center\em#1}}
\newcommand{\email}[1]{\vbox{\center\tt#1}\vspace{5mm}}

\makeatletter\@addtoreset{equation}{section}\makeatother

\begin{document}

\unitlength = .8mm

\begin{titlepage}

\begin{center}

\hfill \\
\hfill \\
\vskip 1cm

\title{On the Ground State Wave Function \\ of Matrix Theory}

\author{Ying-Hsuan Lin and Xi Yin}

\address{Jefferson Physical Laboratory, Harvard University, \\
Cambridge, MA 02138 USA}

\email{yhlin@physics.harvard.edu,
xiyin@fas.harvard.edu}

\end{center}

\abstract{ We propose an explicit construction of the leading terms in the asymptotic expansion of the ground state wave function of BFSS $SU(N)$ matrix quantum mechanics. Our proposal is consistent with the expected factorization property in various limits of the Coulomb branch, and involves a different scaling behavior from previous suggestions. We comment on some possible physical implications. }

\vfill

\end{titlepage}

\eject

\tableofcontents


\section{Introduction}

The matrix theory of Banks-Fischler-Susskind-Shenker \cite{Banks:1996vh, Taylor:2001vb} was formulated by \cite{Balasubramanian:1997kd, Susskind:1998vk, Polchinski:1999br} along the lines of the AdS/CFT correspondence \cite{Maldacena:1997re} as a duality between the 16-supercharge $SU(N)$ gauged matrix quantum mechanics and the decoupling limit of the 0-brane geometry in type IIA string theory, which admits an M-theory lift to an asymptotically null-compactified spacetime. Though the matrix quantum mechanics may appear to be a (deceivingly) simple theory, it has been difficult to extract bulk physics from it. Perturbative computations in matrix theory beyond one-loop suffers from infrared divergences that are regularized through non-perturbative effects \cite{Becker:1997xw}. It is expected that semi-classical gravity in the bulk can only be recovered through strong coupling dynamics at large $N$. Relatively little is known regarding the strong coupling/low energy dynamics of matrix quantum mechanics beyond Monte Carlo simulations. Attempts of analytically understanding the strong coupling dynamics of matrix theory include the use of truncated Schwinger-Dyson equations, with limited success.

Various indirect arguments, as well as a careful computation of the supersymmetric index, indicate that the theory has a unique, normalizable, $SO(9)$ rotationally invariant supersymmetric ground state \cite{Yi:1997eg, Sethi:1997pa, Moore:1998et, Konechby:1998vc, Porrati:1997ej, Sethi:2000zf}. There is a continuum of scattering states above the ground state. It is commonly believed (though not often stated explicitly) that there are no normalizable energy eigenstates of nonzero energy; in other words, all excited energy eigenstates are scattering states. This is consistent with the bulk picture that black holes can decay by radiating D0-branes \cite{Catterall:2009xn}, which are the only particles in the bulk that can escape to infinity. The bulk picture on the other hand also suggests the existence of an exponentially large number of metastable states with exponentially long life time\footnote{This is a peculiar feature of the bulk geometry, in that only the D0-branes can approach asymptotic infinity at a finite cost of energy. It is in contrast to Schwarzschild black holes in flat spacetime whose lifetime scales like a power of its mass.}. These metastable states are the dual description of the microstates of the black hole at finite temperature.

An outstanding question is to describe these metastable states directly in the framework of matrix quantum mechanics. The first step is to understand the structure of the ground state wave function. An asymptotic expansion for the ground state wave function in the $SU(2)$ case has been studied in \cite{Plefka:1997xq,Frohlich:1999zf}, and subsequent proposals for $N\geq 3$ were made in \cite{Hoppe:2000tj, Hasler:2002wt}. In this paper we extend the study of the asymptotic expansion to the general $SU(N)$ matrix theory. We will demonstrate that, first of all, the leading term in the asymptotic ground state wave function is governed by a set of 16 supercharges that describe $N$ or $N-1$ free non-relativistic superparticles on $\mathbb{R}^{9|16}$. This is intuitive from the perspective of effective field theory on the Coulomb branch, though in the EFT approach it was unclear how to carry out a systematic expansion in $1/r$, particularly due to trouble with infrared divergences.

We then propose an explicit form of the leading asymptotic ground state wave function, based on a structure that involves a summation over trees that successively group the $N$ particles. Our proposed form solves the supercharge constraint exactly, and obeys the expected factorization property in various limits on the Coulomb branch of the theory. There is a small ambiguity in our wave function, encoded in a simple set of constant ``two-body coefficients", which are not determined by any simple argument we know of. Our proposal differs from previous suggestions in the $SU(3)$ case \cite{Hoppe:2000tj}; in particular, the overall scaling power with $r$ is different (the proposal of \cite{Hoppe:2000tj} tails off faster at large distances by a factor of $r^{-14}$). We also compute the next-to-leading order correction to the asymptotic wave function, and show how we can go to higher orders.


Let us begin by recalling the Hamiltonian of matrix theory,
\ie
H = {1\over 2} {\rm Tr} \left( P_i^2 - {1\over 2} [X^i, X^j]^2 - \widehat\Theta^T \Gamma^i [X^i, \widehat\Theta] \right),
\fe
where the bosonic and fermionic matrices can be written as $X^i = X^i_A T_A$, $\widehat\Theta_\A = \widehat\Theta_{\A A}T_A$, with $T_A$ the $SU(N)$ generators, normalized by ${\rm Tr} (T_A T_B) = \delta_{AB}$. Here $i=1,2,\cdots,9$ and $\A=1,\cdots,16$ are vector and spinor indices of $SO(9)$. $P_i$ are the canonical momenta conjugate to $X^i$, while $\widehat\Theta_{\alpha A}$ obey canonical anti-commutation relations
\ie
\{ \widehat\Theta_{\A A}, \widehat\Theta_{\B B} \} = \delta_{\A\B} \delta_{AB}.
\fe
Gauging the $SU(N)$ means that we restrict the Hilbert space to consist of $SU(N)$ invariant states.
The 16 supercharges are written as
\ie
Q_\A = {\rm Tr}\left( P_i (\Gamma^i  \widehat\Theta)_\A - {i\over 2} [X^i, X^j] (\Gamma^{ij} \widehat\Theta)_\A \right),
\fe
which obey the supersymmetry algebra up to a gauge rotation
\ie
\{Q_\A, Q_\B\} = 2\delta_{\A\B} H + 2\Gamma^i_{\A\B} X^i_A C_A.
\fe
Here $C_A$ are the operator realization of $SU(N)$ generators,
\ie
C = C_A T_A = -i [X^i, P_i] - {1\over 2} \{ \widehat\Theta_\A, \widehat\Theta_\A \}.
\fe
Our objective is to find the $SO(9)$ invariant ground state wave function annihilated by all $Q_\A$. The idea is to begin with a Born-Oppenheimer-type approximation, by starting at a generic point on the Coulomb branch where the $X^i$'s are close to being commuting with one another, and treat the off-diagonal components as internal degrees of freedom. In the next section we will formulate an expansion of the wave function in powers of $r^{-{3\over 2}}$ where $r$ is essentially the distance between eigenvalues on the Coulomb branch. A (so far) consistent proposal for the leading term in the asymptotic expansion of the ground state is given in section 3. The next-to-leading order correction is computed in section~4, and a systematic way of going to higher orders is presented. We conclude with discussions on the physical implications of our result and some speculations.

\section{The asymptotic expansion}

In this section we explain the method for solving the supersymmetry constraint equations on the wave function based on an asymptotic expansion, closely following the approach of \cite{Frohlich:1999zf} (see also \cite{Hasler:2002wt}).

\subsection{Removing the gauge redundancy}

We are after the $SU(N)$-invariant ground state wave function which is annihilated by the supercharges $Q_\A$, namely
\ie
{\rm Tr} \left\{ {\partial\over \partial X^i} \Gamma^i_{\A\B} \widehat \Theta_\B +{1\over 2}[X^i, X^j] \Gamma^{ij}_{\A\B} \widehat\Theta_\B \right\} \Psi = 0.
\fe
In analyzing the asymptotic form of the wave function, we will expand near a generic point at large distances on the Coulomb branch, and put the bosonic matrices $X^i$ in the form
\ie\label{uxu}
U X^i U^{-1} = \begin{pmatrix} r^i_1 &  0 & & \\ 0 & r^i_2 & & \\ & & \ddots & \\  & &   & r^i_N  \end{pmatrix}
+ \begin{pmatrix} 0 &  q^i_{12} & & \\  (q_{12}^i)^* & 0 & & \\ & & \ddots & \\  & &   & 0  \end{pmatrix}
\fe
for some $SU(N)$ matrix $U$.
We write $\vec r_a = (r^1_a,\cdots, r^9_a)$, $\vec q_{ab} = (q^1_{ab},\cdots, q^9_{ab})$, and work in the regime of large $|\vec r_a - \vec r_b|$ such that $q^i_{ab}$ are very massive. To ensure that this is the case, namely that the $q^i_{ab}$'s are transverse to the valley of the scalar potential, we must choose $U$ in such a way that $\vec q_{ab}\cdot (\vec r_a-\vec r_b)=0$ for all $a,b$. This condition fixes $U$ up to the diagonal $U(1)^{N-1}$ that rotates the phases of $\vec q_{ab}$. We will leave these degrees of freedom in $U$ unfixed. This is acceptable because it still allows us to work in the regime of small $q^i_{ab}$ in the large $r^i_a$ limit. 
Since in this limit $q^i_{ab}$ are described as harmonic oscillators in a potential $|\vec r_a-\vec r_b|^2 (q^i_{ab})^2$, it is convenient to define 
\ie
y^i_{ab} = |\vec r_a-\vec r_b|^{1\over 2} q^i_{ab}
\fe
so that $y^i_{ab} \sim \cO(1)$.


Similarly, we separate $\widehat\Theta_\A$, after the appropriate $SU(N)$ rotation, into diagonal and off-diagonal modes, according to
\ie
U\widehat\Theta_\A U^{-1} = \begin{pmatrix} \theta_{\A1} &  0 & & \\ 0 & \theta_{\A2} & & \\ & & \ddots & \\  & &   & \theta_{\A N}  \end{pmatrix}
+ \begin{pmatrix} 0 &  (\Theta_\A)_{12} & & \\  (\Theta_\A)_{12}^* & 0 & & \\ & & \ddots & \\  & &   & 0  \end{pmatrix}.
\fe
From now the unhatted notation $(\Theta_\A)_{ab}$ will always refer to these off-diagonal components of $U \widehat \Theta_\A U^{-1}$. Note that the overall $SU(N)$ gauge rotation, which acts on both $X^i$ and $\widehat\Theta_\A$, only acts by rotating $U$ and does not act on $(r^i, q^i, \theta_\A, \Theta_\A)$.

The next step is to write ${\partial / \partial X^i}$ in terms of derivatives on $r^i_a$ and $y^i_{ab}$. The details are given in Appendix A, with the result
\ie
\left[ U{\partial \over \partial X^i} U^{-1}\right]_{ba} &= \delta_{ab} {\partial\over \partial r^i_a} + \Pi^{ij}_{ab} {\partial\over \partial q^j_{ab}} 
- {\widehat r_{ab}^i\over |r_{ab}|} \sum_{c\not=a,b}\left( {y^k_{ca}\over |r_{ca}|^{1\over 2} } \Pi^{kj}_{cb}{\partial\over \partial q^j_{cb}} -{y^k_{bc}\over |r_{bc}|^{1\over 2} } \Pi^{kj}_{ac} {\partial\over \partial q^j_{ac}} \right) 
\\
&~~~~+ {\widehat r^i_{ab}\over |\vec r_{ab}|} \left[U {\partial\over \partial U}\right]_{ba}  + {\cal O}(r^{-{5\over 2}}),
\fe
where $r^i_{ab} \equiv r^i_a - r^i_b$, and $\Pi^{ij}_{ab}\equiv \delta^{ij}-\widehat r^i_{ab} \widehat r^j_{ab}$.
Next, we need to change coordinate on the fermions $\widehat\Theta_\A$ into $(\theta_\A,\Theta_\A)$ as well. In doing so, we must make the replacement
\ie
\left[ U{\partial\over \partial U} \right]_{ab} \to R_{ab} + M_{ab},
\fe
where $R_{ab}$ is the overall $SU(N)$ gauge rotation generator that only acts on $U$ but not on $(r^i, q^i, \theta_\A, \Theta_\A)$, and $M_{ab}$ is the $SU(N)$ generator acting on the fermions.\footnote{
Explicitly,
\ie
M_{ab} = {1 \over 2} [ (\Theta_\A)_{ae}, (\Theta_\A)_{ec} ] + ( \theta_{\A a} - \theta_{\A b} ) (\Theta_\B)_{ab}.
\fe
}

Now we can write
\ie
\left[ U{\partial \over \partial X^i} U^{-1}\right]_{ba} &= \delta_{ab} {\partial\over \partial r^i_a} + \Pi^{ij}_{ab} {\partial\over \partial q^j_{ab}} 
- {\widehat r_{ab}^i\over |r_{ab}|} \sum_{c\not=a,b}\left( {y^k_{ca}\over |r_{ca}|^{1\over 2} } \Pi^{kj}_{cb}{\partial\over \partial q^j_{cb}} -{y^k_{bc}\over |r_{bc}|^{1\over 2} } \Pi^{kj}_{ac} {\partial\over \partial q^j_{ac}} \right)  
\\
&~~~~ + {\widehat r^i_{ab}\over |r_{ab}|} (R_{ba} + M_{ba}) + {\cal O}(r^{-{5\over 2}}).
\fe
In the application below, we will take this expression for $\partial/\partial X^i$ to act on an $SU(N)$ invariant wave function, that is, a wave function that is invariant under the $SU(N)$ action simultaneously on the original bosons and fermions $X^i$ and $\Theta_\A$. In the new coordinate system $(U, r^i, q^i, \theta_\A, \Theta_\A)$, it only acts on $U$. The upshot is that $R_{ab}$ annihilates the $SU(N)$ invariant wave function and can be dropped from now, and $U$ will no longer appear explicitly in our computations below.

\subsection{The asymptotic expansion of the supercharge}

After dropping the $R_{ab}$ term and changing variables from $q^i_{ab}$ to $y^i_{ab}$, we can now write the supercharge as an expansion in $r^{-{3\over 2}}$, in the form
\ie
& iQ_\A 
= \sum_{a\not=b} |\vec r_{ab}|^{1\over 2} \left[ \Pi^{ij}_{ab} {\partial\over \partial y^j_{ba}} \Gamma^i_{\A\B} (\Theta_\B)_{ba} + {1\over 2} {\widehat r^i_{ab}} y^j_{ab} \Gamma^{ij}_{\A\B} (\Theta_\B)_{ba} \right] 
\\
& + \sum_a {\partial\over \partial r^i_a} \Gamma^i_{\A\B} \theta_{\B a} +\sum_{a\not=b} \left[ {\widehat r^i_{ab} \over 2|r_{ab}|}y^j_{ab} {\partial\over \partial y^j_{ab}}  \Gamma^i_{\A\B} (\theta_{\B a}-\theta_{\B b}) + \sum_{c\not=a,b}{y^i_{ac} y^j_{cb}\over |r_{ac}|^{1\over 2}|r_{bc}|^{1\over 2}} \Gamma^{ij}_{\A\B} (\Theta_\B)_{ba} \right.
\\
&+  {y^i_{ab} y^j_{ba}\over 2|r_{ab}|} \Gamma^{ij}_{\A\B} (\theta_{\B a} - \theta_{\B b}) -  {\widehat r^i_{ab}\over |r_{ab}|}  \Gamma^i_{\A\B} (\Theta_\B)_{ba} M_{ab} 
\\
& \left. -\sum_{c\not=a,b}\left( {|r_{bc}|^{1\over 2} \over |r_{ac}|^{1\over 2} } y^k_{ca}\Pi^{kj}_{cb}{\partial\over \partial y^j_{cb}} -{|r_{ac}|^{1\over 2} \over |r_{bc}|^{1\over 2} } y^k_{bc}\Pi^{kj}_{ac} {\partial\over \partial y^j_{ac}} \right)  {\widehat r_{ab}^i\over |r_{ab}|} \Gamma^i_{\A\B} (\Theta_\B)_{ab} \right] + {\cal O}(r^{-{5\over 2}}).
\fe
We will write the first line after the equal sign as $Q^0_\A$ and\footnote{Our convention for $Q^n_\A$'s differs from that of the usual supercharge by a factor of $i$.} the next two lines as $Q^1_\A$. $Q^0_\A$ scales like $r^{1\over 2}$ while $Q^1_\A$ scales like $r^{-1}$. The wave function will take the following form
\ie
\Psi = \Psi_0 + \Psi_1 + \Psi_2 + \cdots,
\fe
where $\Psi_n$ scales like $r^{-\kappa - {3\over 2}n}$. Our goal is to determine $\kappa$. Separating the equations according to the scaling degree in $r$, we have a series of equations
\ie
& Q^0_\A \Psi_0 = 0,
\\
& Q^0_\A \Psi_1 + Q^1_\A \Psi_0 = 0,~~~{\rm etc.}
\fe
The first equation is a differential equation in $y^i_{ab}$ only. The solution $\Psi_0$ takes the form
\ie
\Psi_0 = f(\vec r_a) |\psi_0(\widehat r)\rangle_{y,\Theta},
\fe
where $|\psi_0(\widehat r)\rangle_{y,\Theta}$ is the ground state wave function of an $\widehat r_{ab}$-dependent (denoted here collectively by $\widehat r$) supersymmetric harmonic oscillator in the off-diagonal $(y, \Theta)$ sector, obeying
\ie
Q_\A^0 |\psi_0(\hat r)\rangle_{y, \Theta} \equiv \sum_{a\not=b} |r_{ab}|^{1\over 2} \left[  \Pi^{ij}_{ab}{\partial\over \partial y^j_{ba}} \Gamma^i_{\A\B} (\Theta_\B)_{ba} + {1\over 2} {\widehat r^i_{ab} } y^j_{ab} \Gamma^{ij}_{\A\B} (\Theta_\B)_{ba} \right] |\psi_0(\hat r)\rangle_{y, \Theta} = 0.
\label{q0p}
\fe
$f(\vec r_a) $ is a so far undetermined wave function that has some overall scaling $r^{-\kappa}$, and includes the fermionic wave function in the diagonal $\theta$ sector. 

The next key step is to consider a projection $P_0$ onto the zero-eigenspace of $Q^0_\A$.
Since $iQ_\A^0$ is Hermitian, any state of the form $Q_\A^0 \Psi_1$ must be orthogonal to the zero-eigenspace of $Q_\A^0$, and is thus annihilated by $P_0$. Consequently, the next-to-leading order equation in the asymptotic expansion implies
\ie
P_0 Q^1_\A \Psi_0 = 0.
\fe
Since $Q^1_\A$ involves an $r$-derivative, this equation will provide nontrivial constraints on $f(\vec r_a)$.

\subsection{Reducing to the Cartan wave function}

$Q^0_\A$ has an anti-commutator of the form
\ie
\{ Q_\A^0, Q_\B^0 \} 
& = \delta_{\A\B} \sum_{a\not=b} |r_{ab}|\left[ \Pi^{ij}_{ab} {\partial \over\partial y^i_{ab}}{\partial \over\partial y^j_{ba}} - {1\over 4} y_{ab}\cdot y_{ba}  - {1\over 2} \widehat r^k_{ab} \Gamma^k_{\C\D} (\Theta_\C)_{ab} (\Theta_\D)_{ba} \right]
+ \Gamma^k_{\A\B} {\cal M}_k.
\fe
For each pair $a,b$, consider the matrix $\widehat r^k_{ab} \Gamma^k_{\A\B}$ that acts on $SO(9)$ spinors. This matrix has eight $+1$ eigenvalues and eight $-1$ eigenvalues. Let $\Pi^\pm_{ab}$ be the projection operators onto the positive and negative spinor eigenspaces of $\widehat r_{ab}^i \Gamma^i$.
By definition, $\Pi^\pm_{ab} = \Pi^\mp_{ba}$.

Given a fixed pair $a,b$, let $|F_{ab}(\widehat r_{ab})\rangle$ be a unit norm state in the $\Theta_{ab}$ sector, that is annihilated by $(\Theta^-_{ab})_\A\equiv (\Pi^-_{ab})_{\A\B}(\Theta_\B)_{ab}$ and $(\Theta^-_{ba})_\A\equiv(\Pi^-_{ba})_{\A\B}(\Theta_\B)_{ba} = (\Pi^+_{ab})_{\A\B}(\Theta_\B)_{ab}^*$ for all $\A$, and is invariant under simultaneous $SO(9)$ rotations on $\widehat r_{ab}$, $\Theta_{ab}$ {\it and} $\Theta_{ba}$. Such a state is unique up an overall ($\widehat r$-independent) phase. We will write $|F(\widehat r)\rangle = \bigotimes_{a<b}|F_{ab}(\widehat r_{ab})\rangle$ for such a zeroth-order fermion ground state in the entire off-diagonal $\Theta$ sector (again, the notation here is such that $\widehat r$ stands collectively for the set of all $\widehat r_{ab}$'s). We can then construct $|\psi_0(\widehat r)\rangle$ by combinging $|F(\widehat r)\rangle$ with the harmonic oscillator ground state wave function for the $y^i_{ab}$'s,
\ie
|\psi_0(\widehat r)\rangle = e^{-{1\over 4} \sum_{a\not=b}|y_{ab}|^2} |F(\widehat r)\rangle.
\fe
There are $8N(N-1)$ independent $y^i_{ab}$'s, and the ground state energy of the harmonic oscillator precisely cancels with the fermionic contribution in the coefficient of $\delta_{\A\B}$. One can verify that $|\psi_0(\widehat r)\rangle$ is annihilated by ${\cal M}_k$ as well.

Now we can write
\ie
\Psi_0 =  e^{-{1\over 4} \sum_{a\not=b}|y_{ab}|^2} \sum_s f_s(\vec r_a)|s\rangle \otimes |F(\widehat r)\rangle,
\fe
for a set of functions $f_s(\vec r_a)$, where $s$ labels states in the Clifford module of the $16(N-1)$ diagonal $\theta_{\A a}$'s ($s= 1,\cdots, 2^{8(N-1)}$). Let us inspect the action of
\ie
Q^1_\A &= \sum_a {\partial\over \partial r^i_a} \Gamma^i_{\A\B} \theta_{\B a} +\sum_{a\not=b} \left[ {\widehat r^i_{ab} \over 2|r_{ab}|}y^j_{ab} {\partial\over \partial y^j_{ab}}  \Gamma^i_{\A\B} (\theta_{\B a}-\theta_{\B b}) + \sum_{c\not=a,b} {y^i_{ac} y^j_{cb}\over |r_{ac}|^{1\over 2}|r_{bc}|^{1\over 2}}\Gamma^{ij}_{\A\B}  (\Theta_\B)_{ba} \right.
\\
&~~~~ + {y^i_{ab} y^j_{ba}\over 2| r_{ab}|} \Gamma^{ij}_{\A\B} (\theta_{\B a} - \theta_{\B b})
- {\widehat r^i_{ab}\over | r_{ab}|}  \Gamma^i_{\A\B} (\Theta_\B)_{ba} M_{ab}
\\
&~~~~ \left. - \sum_{c\not=a,b}\left( {|r_{bc}|^{1\over 2} \over |r_{ac}|^{1\over 2} } y^k_{ca}\Pi^{kj}_{cb}{\partial\over \partial y^j_{cb}} -{|r_{ac}|^{1\over 2} \over |r_{bc}|^{1\over 2} } y^k_{bc}\Pi^{kj}_{ac} {\partial\over \partial y^j_{ac}} \right)  {\widehat r_{ab}^i\over |r_{ab}|} \Gamma^i_{\A\B} (\Theta_\B)_{ab} \right]
\fe
on $\Psi_0$. Keep in mind that $\partial/\partial r^i_a$ which appears in $Q^1_\A$ acts not only on the functions $f_s(\vec r_a)$ but on $|F(\widehat r)\rangle$ as well. 

Under the projection $P_0$, we can replace $y_{ab}\cdot \partial_{y_{ab}}$ and $y^i_{ac} y^j_{cb}$ in $Q^1_\A$ by their expectation values in the harmonic oscillator ground state wave function $e^{-{1\over 4} \sum_{a,b} |y_{ab}|^2} = e^{-{1\over 2}\sum_{a<b} |y_{ab}|^2}$. 
Furthermore, any term that involves the product of an odd number of $\Theta$'s when acting on $\Psi_0$ cannot preserve the fermion ground state in the $\Theta$ sector, and the result will be annihilated by $P_0$. Note that the projector $P_0$ does not touch the $\theta_{\A a}$ degrees of freedom.
%
Let us define $(\Theta^\pm_\A)_{ab} \equiv \Pi^+_{ab} (\Theta_\A)_{ab}$. All states that survive the $P_0$ projection are annihilated by $\Theta_\A^-$, while any state obtained by acting with $\Theta_\A^+$ is killed by $P_0$. 
Using the relation
\ie
& P_0 (\Theta_\B)_{ba} M_{ab} \Psi_0 
= P_0 [(\Theta_\B)_{ba}^-, M_{ab}] \Psi_0
= -(\Pi^+_{ab}(\theta_a-\theta_b))_{\B} \Psi_0,
\fe
we can replace $Q^1_\A$ by a simplified operator
\ie\label{qac}
\widetilde Q^1_\A &=  \sum_a {\partial\over \partial r^i_a} \Gamma^i_{\A\B} \theta_{\B a}  - \sum_{a\not=b} {2\over |r_{ab}|} \widehat r_{ab}^i\Gamma^i_{\A\B} (\theta_{\B a}-\theta_{\B b})+ \sum_{a\not=b} {1\over |r_{ab}|} (\Pi^+_{ab})_{\A\B} (\theta_{\B a}-\theta_{\B b})
\\
&= \sum_{a}  {\partial\over \partial r^i_a} \Gamma^i_{\A\B} \theta_{\B a}  - \sum_{a\not=b}  {3\over |r_{ab}|} \widehat r_{ab}^i\Gamma^i_{\A\B} \theta_{\B a},
\fe
in the sense that
\ie
P_0 Q^1_\A \Psi_0 = P_0 \widetilde Q^1_\A \Psi_0.
\fe
Furthemore, the $r^i_a$ dependence of $\Psi_0$ may be expressed as dependence on $|r_{ab}|$ and $\widehat r_{ab}$. Under a variation $\delta r^i_a$, we have
\ie
& \delta |r_{ab}| = {\widehat r_{ab}}\cdot (\delta \vec r_a - \delta \vec r_b),
\\
& \delta \widehat r_{ab} = {\delta \vec r_{ab} - \widehat r_{ab} (\widehat r_{ab}\cdot\delta\vec r_{ab}) \over |r_{ab}|} .
\fe
Thus we can write
\ie
{\partial\over \partial r^i_a} 
= \sum_{b\not=a}\left({\widehat r_{ab}^i} {\partial\over \partial |r_{ab}|}
+ {\widehat r_{ab}^j\over |r_{ab}|} R^{ji}_{ab} \right), 
\fe
where $R^{ij}_{ab}$ is the generator of $SO(9)$ rotation on $\widehat r_{ab}$ for each pair $a,b$. Note that it does not act on the fermions, by definition. 
In the $\Theta$ sector, the zeroth order ground state wave function $|F(\widehat r)\rangle$ by construction is invariant under the $SO(9)$ rotation on $\widehat r_{ab}$, $\Theta_{ab}$, {\it and} $\Theta_{ba}$. Let us denote by $F^{ij}_{ab}$ the $SO(9)$ rotation generator on $\Theta_{ab}$ and $\Theta_{ba}$, namely
\ie
F^{ij}_{ab} = {1\over 4}(\Theta_{ab} \Gamma^{ij} \Theta_{ba}).
\fe
Thus when acting on $|F(\widehat r)\rangle$ with $R^{ij}_{ab}$, we can replace $R^{ij}_{ab}$ by $-F^{ij}_{ab}$. Note that $F^{ij}_{ab}|F(\widehat r)\rangle = {1\over 4}(\Theta_{ab}^+ \Gamma^{ij} \Theta_{ba}^+)|F(\widehat r)\rangle$, and is thus annihilated by the projector $P_0$. In other words, we can ignore the $\widehat r_{ab}$-dependence of $|F(\widehat r)\rangle$ in computing $P_0\widetilde Q^1_\A \Psi_0$. For this purpose, we might as well replace $\widetilde Q^1_\A$ by an operator\footnote{By a slight abuse of notation we will still denote this operator by $\widetilde Q^1_\A$.} of the same form as (\ref{qac}), but now acting entirely on the ``Cartan wave function" 
\ie
\Psi^C_0 = \sum_s f_s(\vec r_a)|s\rangle
\fe
that is just in the $(r, \theta)$ sector.
Now the projector $P_0$ is no longer needed; the equation $P_0 Q^1_\A \Psi_0=0$ simply reduces to
\ie
\widetilde Q^1_\A \Psi_0^C = 0.
\fe

\subsection{Treating the Cartan fermions}

In the simplest $SU(2)$ case, the indices $a,b$ take values 1 and 2 (and $\vec r_2 = -\vec r_1$). 
There are 16 $\theta_\A$'s, giving rise to $2^8=256$ states in the $\theta$ sector. With respect to the $SO(9)$ rotation on the $\theta_\A$'s, these 256 states branch into 
\ie
{\bf 44}\oplus {\bf 84}\oplus {\bf 128}.
\fe
Here the ${\bf 44}$ is the traceless symmetric tensor representation of $SO(9)$. The other two irreducible representations of $SO(9)$ cannot form a singlet by tensoring with a power of the vector representation (coming from $\widehat r$). 
The fermion part of the $SO(9)$ invariant ground state wave function, $|s\rangle$, must thus be constructed from the ${\bf 44}$. Such a state is unique up to the overall factor, namely, it is $|\widehat r \widehat r \rangle \equiv \widehat r^i \widehat r^j |s_{ij}\rangle$, where $|s_{ij}\rangle$ is a basis for the ${\bf 44}$. 
The $SO(9)$ invariance of the wave function allows us to replace $R^{ij}$ by $- {1\over 4}(\theta\Gamma^{ij}\theta)$ that rotates $\theta$ instead of $\vec r$. One can show that
\ie
\widehat r^j (\Gamma^i\theta)_\A(\theta\Gamma^{ij}\theta)|\widehat r\widehat r\rangle = 36 \widehat r^i (\Gamma^i \theta)_\A |\widehat r\widehat r\rangle.
\fe 
One then finds that $\widetilde Q^1_\A \Psi_0^C=0$ is solved by $\Psi_0^C = r^{-6} |\widehat r\widehat r\rangle$.

The case of general $SU(N)$ gauge group will be treated in the next section.
Note that the integration measure for our wave function $\Psi$ at large $r$ takes the form\footnote{
Here $r^2_{ab}$ come from the gauge-fixing, and $r^{-4N(N-1)}$ comes from the change of variables from $q$ to $y$.
}
\ie
\int \prod_{a=1}^{N-1} d^9 \vec r_a\, \left( \prod_{a<b} r_{ab}^2 \right) \, r^{-4N(N-1)} \int \prod_{a\not=b} d^9\vec y_{ab} ~ \delta(\vec y_{ab}\cdot\widehat r_{ab}).
\fe
If the leading asymptotic wave function $\Psi_0$ has an overall scaling $r^{-\kappa}$, normalizability then demands $\kappa > -{3\over 2}(N-3)(N-1)$.

\section{The leading ground state wave function}

\subsection{Reducing to free superparticles}

We are seeking an $S_N\times SO(9)$ invariant Cartan wave function $\Psi_0^C$ that is annihilated by
\ie
\widetilde Q^1_\A 
&= \sum_a \left( {\partial\over \partial r^i_a}  - \sum_{b\not=a} {3\over r_{ab}^2} r_{ab}^i\right) \Gamma^i_{\A\B} \theta_{\B a}.
\fe
It is convenient to define
\ie
\Psi^{new} \equiv \prod_{a<b}|r_{ab}|^{-3} \Psi_0^C.
\fe
Then the equation for $\Psi^{new}$ becomes simply $Q_\A^{new}\Psi^{new}=0$, where $Q_\A^{new}$ take the form of the supercharges for a set of free superparticles,
\ie
Q_\A^{new} = \sum_a {\partial\over \partial r^i_a} \Gamma^i_{\A\B} \theta_{\B a}.
\fe
We immediately learn that $\Psi^{new}$ takes the form
\ie
\Psi^{new} = \sum_s F_s(r_a^i) |s\rangle,
\fe
where $F_s(r_a^i)$ for each internal fermion state $|s\rangle$ is a harmonic function on $\mathbb{R}^{9(N-1)}$.
Indeed, in the $SU(2)$ case, $\Psi^{new}=r^{-9} \widehat r^i \widehat r^j |s_{ij}\rangle\propto\partial_i \partial_j r^{-7}|s_{ij}\rangle$ is of such form.


\subsection{The $SU(N)$ proposal}
 
So far we have been writing the supercharges and the Hamiltonian as if we were dealing with the $U(N)$ theory. In dealing with the $SU(N)$ matrix theory, we need to factor out the center of mass degrees of freedom. This is straightforward in the bosonic sector: the wave function when viewed as a function of $\vec x_1, \cdots, \vec x_N$ is taken to be invariant under the overall translation $\vec P =\sum_{a=1}^N \vec p_a$. Care must be taken in the fermion sector, however, since we have quantized the $\theta_{\A a}$ independently, with
\ie
\{ \theta_{\A a} , \theta_{\B b} \} = \delta_{ab}\delta_{\A\B}.
\fe
We should factor out $\overline{\theta} = (\theta_1+\theta_2+\cdots+\theta_N)/N$, and only work with the combinations of $\theta$'s (for instance, $\theta_a - \overline\theta$) that anti-commute with $\overline\theta$. In the expression for the supercharge $Q_\A$ in terms of $r^i_a, q^i_{ab}, \theta_{\A a}, (\Theta_\A)_{ab}$, the only term that involves the center of mass position and fermionic coordinate $\overline\theta$ is $\sum_{a=1}^N p^i_a \Gamma^i\theta_a$, where $p^i_a = -i \partial/\partial r^i_a$. In passing to the $SU(N)$ system, we can separate
\ie
\sum_{a=1}^N p^i_a \Gamma^i\theta_a = P^i \Gamma^i\overline\theta + \sum_{a=1}^N \left(p^i_a-{1\over N} P^i\right) \Gamma^i(\theta_a-\overline\theta),
\fe
and simply drop the first term $P^i\Gamma^i\overline\theta$, since $P^i$ and $\overline\theta$ commute with the remaining terms of the supercharge. The ground state wave function will depend on the relative bosonic coordinates $\vec x_a - \vec x_b$, and its fermionic component may be constructed as an element of the Clifford module coming from $\theta_a - \overline\theta$. Be aware that $\theta_a - \overline\theta$ do not anti-commuate with $\theta_b-\overline\theta$ for $a\not=b$. Rather, we have
\ie
\left\{ \theta_a - \overline\theta, \theta_b - \overline\theta \right\} = \delta_{ab} - {1\over N}.
\fe
One can in principle go to a basis in which the anti-commutators become diagonal, and quantize the theory using that basis.  However, such a basis is rather inconvenient to work with.  Below we will employ a different approach.
 
Though the problem of finding $\Psi_0$ is reduced to the free problem of finding $\Psi^C$ or $\Psi^{new}$, this problem doesn't have a unique solution in the general $SU(N)$ case, even after imposing $S_N\times SO(9)$ invariance. It is possible that there are more constraints coming from the smoothness of the wave function at small $r^i_{ab}$ when all order corrections are included. For now, we will constrain $\Psi_0$ further by some physical intuition.
Namely, we expect that in a limit on the Coulomb branch where $(r^i_a, \theta_\A)$ are separated into two clusters centered at $(x^i, \theta_\A)$ and $(y^i, \eta_\A)$, and the $SU(N)$ broken into $SU(M)\times SU(N)$, $\Psi^{new}$ should be approximately proportional to the $SU(2)$ wave function in the relative bosonic and fermionic coordinates $(x^i-y^i, \theta_\A - \eta_\A)$. Motivated by this, we now make a proposal for $\Psi_0$ (or equivalently for $\Psi^{new}$) which will be an exact solution of $P_0 Q^1_\A \Psi_0=0$, and satisfies this factorization criterion.

We will in fact define recursively a weighted $n$-body asymptotic wave function,
\ie
\Psi^{(n)}_{k_1,k_2,\cdots,k_n}(\vec r_1, \widehat\theta_{1\A}; \vec r_2, \widehat\theta_{2\A}; \cdots; \vec r_n, \widehat\theta_{n\A}).
\fe
Here $k_a$ are a set of positive integers. By writing $\widehat\theta_{\A a}$ in the argument, we simply mean that the fermionic component of the wave function is built by quantization of $\widehat\theta_{\A a}$ according to their appropriate anti-commutators. We will see in the construction below that $\widehat\theta_{\A a}$ obey the anti-commutation relations
\ie
\{\widehat\theta_{\A a}, \widehat\theta_{b\B}\} ={1\over k_a}\delta_{ab}\delta_{\A\B}.
\fe
In fact, by construction $\Psi_{k_1,\cdots, k_n}$ will be a function of the relative positions $\vec r_a - \vec r_b$ only, and its fermion component will be built out of $\widehat\theta_{\A a}-\widehat\theta_{\A b}$ only.

First of all, we define a two-body wave function,
\ie
\Psi^{(2)}_{k_1,k_2}(\vec r_1, \widehat\theta_{\A1}; \vec r_2, \widehat\theta_{\A2}) = C_{k_1, k_2} \Psi^{new}_{SU(2)} \left({\vec r_1 - \vec r_2\over \sqrt{k_1^{-1} + k_2^{-1} }}, {\widehat\theta_{\A1} - \widehat\theta_{\A2} \over \sqrt{k_1^{-1} + k_2^{-1} }}\right).
\fe
Here $\Psi_{SU(2)}^{new}(\vec r, \theta)$ is as in the $SU(2)$ case,
\ie
\Psi_{SU(2)}^{new}(\vec r, \theta) = \sum_{i,j=1}^9 \partial_i \partial_j |\vec r|^{-7} |s_{ij}\rangle_{\theta}.
\fe
$C_{k_1,k_2}=C_{k_2,k_1}$ is a normalization constant that may depend on $k_1, k_2$, which is so far undetermined. Note that the two-body wave function factor is invariant under exchanging the two bodies ($\vec r\to -\vec r,\theta\to -\theta$).

Now we define the recursive relation between the $n$-body wave function and the $(n-1)$-body wave function
\ie
&\Psi^{(n)}_{k_1,k_2,\cdots,k_n}(\vec r_1, \widehat\theta_{1\A}; \vec r_2, \widehat\theta_{2\A}; \cdots; \vec r_n, \widehat\theta_{n\A}) = \sum_{1\leq i<j\leq n} C_{k_i,k_j} \Psi^{new}_{SU(2)}\left({\vec r_i - \vec r_j\over \sqrt{k_i^{-1} + k_j^{-1} }}, {\widehat\theta_i - \widehat\theta_j\over \sqrt{k_i^{-1} + k_j^{-1}}}\right) 
\\
&~ \times \Psi^{(n-1)}_{k_i+k_j, k_1,\cdots,\cancel{k_i},\cdots,\cancel{k_j},\cdots,k_n}
\left({k_ir_i + k_jr_j \over k_i+k_j}, {k_i\widehat\theta_i + k_j\widehat\theta_j\over k_i+k_j}; r_1, \widehat\theta_1;\cdots; \xcancel{r_i}, \xcancel{\widehat\theta_i}; \cdots; \xcancel{r_j}, \xcancel{\widehat\theta_j}; \cdots; r_n, \widehat\theta_n\right).
\fe
Note that by our construction, $ {\widehat\theta_i - \widehat\theta_j\over \sqrt{k_i^{-1} + k_j^{-1}}}$ anti-commutes with ${k_i\widehat\theta_i + k_j\widehat\theta_j\over k_i+k_j}$ and with all other $\widehat\theta_k$, $k\not=i,j$.

It is then straightforward to verify that
\ie
\Psi^{new} = \Psi^{(N)}_{1,1,\cdots,1}(\vec r_1, \theta_{1\A}; \vec r_2, \theta_{2\A}; \cdots; \vec r_N, \theta_{N\A})
\fe
is an exact solution for the asymptotic ground state Cartan wave function, namely the corresponding $\Psi_0^C$ is annihilated by $\widetilde Q^1_\A$.\footnote{This is easily seen from the simple identity under the change of variables 
\ie
& r^- = {r_1-r_2\over \sqrt{k_1^{-1} + k_2^{-1}}}, ~~~~ r^+ = {k_1r_1+k_2r_2\over k_1+k_2},
\\
& \theta^- = {\widehat\theta_1-\widehat\theta_2\over \sqrt{k_1^{-1} + k_2^{-1}}}, ~~~~ \theta^+ = {k_1\widehat\theta_1+k_2\widehat\theta_2\over k_1+k_2},
\fe
that
\ie
\widehat\theta_1 {\partial\over \partial r_1}  + \widehat\theta_2 {\partial\over \partial r_2} =\theta^- {\partial\over \partial r^-}  + \theta^+ {\partial\over \partial r^+}. 
\fe
The normalization factors are needed in order to preserve the desired normalization of the anti-commutators of $\widehat\theta$'s.}

The proposed $\Psi^{new}$ is also manifestly invariant under the permutation (Weyl group action) by $S_N$, and is $SO(9)$ rotationally invariant. And it satisfies the factorization property in various limits of the Coulomb branch with the symmetry breaking pattern $SU(N)\to SU(k)\times SU(N-k)$. To see the latter, consider the limit where say a cluster $\vec r_1, \cdots,\vec r_k \sim \vec R_1$ are far separated from $\vec r_{k+1},\cdots,\vec r_N\sim \vec R_2$. In this limit $\Psi^{new}$ is dominated by 
\ie
\Psi^{new} \longrightarrow ~ & C_{k,N-k}\Psi^{new}_{SU(2)}\left(\sqrt{k(N-k)\over N}\,(\vec R_1-\vec R_2), \sqrt{N-k\over k N}\,(\widehat\theta_1+\cdots+\widehat\theta_k)\right.
\\
&\hspace{1in} \left.- \sqrt{k\over (N-k)N}\,(\widehat\theta_{k+1}+\cdots+\widehat\theta_N) \right) \\
& \times \Psi^{(k)}_{1,\cdots,1}(r_1,\widehat\theta_1;\cdots;r_k,\widehat\theta_k)\,
\Psi^{(N-k)}_{1,\cdots,1}(r_{k+1},\widehat\theta_{k+1};\cdots;r_N,\widehat\theta_N),
\fe
which scales like $|\vec R_1-\vec R_2|^{-9}$ at large separations between the two clusters. The contributions from other terms in the recursive sum die off like $|\vec R_1-\vec R_2|^{-18}$ or faster in this limit.

$\Psi^{new}$ may also be expressed as a summation over all trees that join the $N$ particles, the product of two-body wave functions associated with each bifurcation of the tree, weighed by the coefficient $\prod_{bifurcation} C_{k_i,k_j}$.

Note that the asymptotic wave function $\Psi_0$ is not normalizable, obviously, since it is homogeneous under the simultaneous rescaling of all $\vec r_a$. We don't have an a priori argument to fix the coefficients $C_{k_1, k_2}$. It is perhaps tempting to suggest that $C_{k_1,k_2}=1$ for all $k_1,k_2$, but this need not be the case. Even though the full two-body wave function has a natural normalization, $\Psi^{(2)}_{k_1,k_2}$ only captures its tail at large distances. 

This proposal would easily answer the question of the overall scaling exponent in $r$. $\Psi^{new}$ scales like $r^{-9(N-1)}$, and therefore
\ie
\kappa = -{3\over 2}N(N-1) + 9(N-1).
\fe
The power of convergence in the integration of the squared wave function at large $r$ is then $r^{-9(N-1)}$. This is different from the previous proposal of \cite{Hoppe:2000tj} in the $SU(3)$ case, for instance. The ansatz of \cite{Hoppe:2000tj} is constructed by taking an $\vec r_a$-independent $SO(9)$ singlet fermion wave function, multiplied by the scalar harmonic function $r^{-9(N-1)+2}$, and then acted on by all 16 free supercharges $Q^{new}_\A$. The resulting wave function falls off faster than our proposal by a factor of $r^{-14}$ at large distances.


\section{Going to higher orders}

\subsection{The general structure}

Now that we have found a solution for $\Psi_0$ that obeys
\ie
P_0 Q^1_\A \Psi_0 = 0,
\fe
we can then determine $\Psi_1$ as
\ie\label{psione}
\Psi_1 = {1\over 16 H^0} Q^0_\A Q^1_\A \Psi_0 + {\cal K}_1,
\fe
where $-16H^0 = Q^0_\A Q^0_\A$ (this comes from $\{Q^0_\A, Q^0_\B\} = -2H^0 \delta_{\A\B} + \Gamma^k_{\A\B} {\cal M}_k^0$), and ${\cal K}_1$ is a yet to be determined wave function in the kernel of $H^0$ (or of the $Q^0_\A$'s). 
It follows from the Jacobi identity on the $Q_\A$'s expanded to first order that (\ref{psione}) indeed solves the equation $Q^0_\B\Psi_1+Q^1_\B\Psi_0=0$. 

The next equation in the $r^{-{3\over 2}}$ expansion is
\ie
Q^0_\A \Psi_2 + Q^1_\A \Psi_1 + Q^2_\A \Psi_0 = 0.
\fe
Not knowing $\Psi_2$, we can again project by $P_0$, and consider 
\ie
P_0 Q^1_\A \Psi_1 + P_0 Q_\A^2 \Psi_0 = 0.
\fe
This may be expressed as an equation for ${\cal K}_1$, 
\ie\label{kkeq}
P_0 Q^1_\A {\cal K}_1 = -P_0\left( Q^1_\A{1\over 16H^0} Q^0_\B Q^1_\B + Q^2_\A \right) \Psi_0.
\fe
The situation here is similar to the equations for $\Psi_0$. We could demand ${\cal K}_1$ to be a Cartan wave function tensored with $|\psi_0(\widehat r)\rangle$ (the unique ground state of $H^0$ in the $(y,\Theta)$ sector), and then try to solve a Dirac-like equation for free superparticles, but now with a source term.

In fact, the RHS of (\ref{kkeq}) vanishes. This can be seen by inspecting the general structure of the RHS of (\ref{kkeq}). $Q^0_\B Q^1_\B\Psi_0$ is a linear combination of states in the $(y,\Theta)$ sector that has $H^0$ eigenvalues ${1\over 2}|r_{ab}|$, $|r_{ab}|$, or ${3\over 2}|r_{ab}|$. It is straightforward to compute $(H^0)^{-1} Q_\B^0 Q^1_\B \Psi_0$ explicitly, which we defer to the next subsection. When we act on it further with $P_0 Q^1_\A$, only the $(y,\Theta)$-sector lowering operators in $Q^1_\A$ contribute. In the end, we can write $P_0 Q^1_\A(H^0)^{-1} Q_\B^0 Q^1_\B \Psi_0$ in a way such that no $\vec r_a$-derivatives are taken on $\Psi_0$. Now $Q^0_\A$ changes the total level in the $y$-sector by an odd amount, while $Q^1_\A$ contains only terms that change the total $y$-level by an even amount. Thus $Q^1_\A(H^0)^{-1} Q_\B^0 Q^1_\B \Psi_0$ must be excited in the $y$-sector and is annihilated by $P_0$.

As for the term $P_0 Q^2_\A \Psi_0$ on the RHS of (\ref{kkeq}), once again we need only consider the terms in $Q^2_\A$ that leave the $(y,\Theta)$ sector in its ground state. It is not hard to see that $Q^2_\A$ has the schematic form $\theta y \partial_r + \theta y^2 \partial_y + \Theta y \partial_y + y \Theta^3 + y \theta \Theta^2 + y^3 \partial_y \Theta$. The last term comes from expanding $\partial q^j_{ab}/\partial X^i$ to one order higher than what is computed explicitly in Appendix A. We don't need its explicit form nonetheless. None of these terms could keep both $y$ and $\Theta$ sectors in their ground states. We conclude that $P_0 Q^2_\A \Psi_0=0$.

%
%

So in the end ${\cal K}_1$ obeys exactly the same equations as that of $\Psi_0$, and can be set to zero.\footnote{More precisely, it can be absorbed into $\Psi_0$, which isn't a priori homogeneous. Though our proposal for $\Psi_0$ is homogeneous with respect to the simultaneous rescaling of all $\vec r_a$, in principle there could be corrections of subleading power in $r$, for instance the type of solution considered in \cite{Hoppe:2000tj}. } 

\subsection{Solving for $\Psi_1$}

The next-to-leading order asymptotic wave function $\Psi_1$ is thus given by ${1\over 16} (H^0)^{-1} Q^0_\A Q^1_\A \Psi_0$. We can put $Q^1_\A \Psi_0 = (1-P_0) Q^1_\A \Psi_0$ into the form
\ie
&Q^1_\A \Psi_0 = \sum_{a\not=b} \left[ {\widehat r_{ab}^j\over 4|r_{ab}|} (\Theta_{ab}^+\Gamma^{ij}\Theta_{ba}^+)\Gamma^i_{\A\B}\theta_{\B a} + {\widehat r^i_{ab}\over 2|r_{ab}|} \left( y^j_{ab} {\partial\over \partial y^j_{ab}} + 4\right) \Gamma^i_{\A\B} (\theta_{\B a} - \theta_{\B b})  \right.
\\
&~~~~ +\sum_{c\not=a,b} {y^i_{ac} y^j_{cb}\over | r_{ac}|^{1\over 2} | r_{bc}|^{1\over 2} }\Gamma^{ij}_{\A\B} (\Theta_\B^+)_{ba} 
+  {y^i_{ab} y^j_{ba}\over 2| r_{ab}|} \Gamma^{ij}_{\A\B} (\theta_{\B a} - \theta_{\B b}) 
- {\widehat r_{ab}^i \over |r_{ab}|} (1-P_0)\Gamma^i_{\A\B}(\Theta_\B)_{ba}M_{ab}
\\
&~~~~  \left. -\sum_{c\not=a,b}\left( {|r_{bc}|^{1\over 2} \over |r_{ac}|^{1\over 2} } y^k_{ca}\Pi^{kj}_{cb}{\partial\over \partial y^j_{cb}} -{|r_{ac}|^{1\over 2} \over |r_{bc}|^{1\over 2} } y^k_{bc}\Pi^{kj}_{ac} {\partial\over \partial y^j_{ac}} \right)  {\widehat r_{ab}^i\over |r_{ab}|} \Gamma^i_{\A\B} (\Theta_\B^+)_{ab} 
  \right]\Psi_0
\fe
It is straightforward though tedious to compute $Q^0_\A Q^1_\A \Psi_0$. By inspecting the excitation levels in the $(y, \Theta)$-sector, we can easily act $(H^0)^{-1}$ on it and obtain, after some simplification,
\ie\label{hqq}
&\Psi_1 = -{5\over 8} \sum_{a\neq b} {1 \over |r_{ab}|^{3\over 2}} (\Theta_{ba}^+\, {\slash\!\!\! y}_{ab} (\theta_a-\theta_b)) \Psi_0 +\sum_{a\neq b} \sum_{c \neq a, b} {1\over |r_{ab}|+ |r_{ac}|+ |r_{bc}|}
\left[ {15\over 8} {(\Theta_{bc}^+ \, {\slash\!\!\! y}_{ca} \Theta_{ab}^+)\over|r_{ac}|^{1\over 2}} \right.
\\
&  
\left. ~~~~~~~~ +{1\over 16}\left( {1\over |r_{bc}|} - {1\over |r_{ab}|} \right) {(\vec r_{ab}\cdot \vec y_{ca}) (\Theta^+_{ab}\Theta^+_{bc}) \over | r_{ac}|^{1\over 2} }  -2{(\vec r_{cb}\cdot \vec y_{ac}) (\vec y_{cb}\cdot\vec y_{ba}) \over |r_{ab}|^{1\over 2} |r_{ac}|^{1\over 2}|r_{bc}|^{1\over 2}} \right] \Psi_0.
\fe

\subsection{Higher orders in the $r^{-{3\over 2}}$ expansion }

While the first order correction $\Psi_1$ is determined algebraically from $\Psi_0$, this is a priori not the case at higher orders. For instance, in order to solve for $\Psi_2$, we need to consider the following two equations. The first one is
\ie
& Q^0_\A \Psi_2 + Q^1_\A \Psi_1 + Q^2_\A \Psi_0 = 0
\\
& \Rightarrow~ \Psi_2 = {1\over 16H^0} \left( Q^0_\A Q^1_\A \Psi_1 + Q^0_\A Q^2_\A \Psi_0 \right)
+{\cal K}_2,
\fe
where ${\cal K}_2$ obeys $Q^0_\A {\cal K}_2=0$. Here we are separating $\Psi_2$ into a piece that involves excited states in the off-diagonal $(y,\Theta)$ sector, and a piece ${\cal K}_2$ that involves only the ground state in the off-diagonal sector. The second equation we need to consider is
\ie
& Q^0_\A \Psi_3 + Q^1_\A \Psi_2 + Q^2_\A \Psi_1 + Q^3_\A \Psi_0 = 0
\\
& \Rightarrow~ P_0Q^1_\A \Psi_2 + P_0Q^2_\A \Psi_1 + P_0Q^3_\A \Psi_0 = 0.
\fe
${\cal K}_2$ can now be determined from
\ie\label{ktwo}
& P_0 Q^1_\A {\cal K}_2 = - P_0 Q^1_\A  {1\over 16 H^0} \left( Q^0_\B Q^1_\B \Psi_1 + Q^0_\B Q^2_\B \Psi_0 \right)  - P_0\left( Q^2_\A \Psi_1 +Q^3_\A \Psi_0\right) .
\fe
The RHS of (\ref{ktwo}) appears to be nontrivial, and now we need to solve a Dirac-like equation for the wave function of $N-1$ superparticles with a source. Note that while we demand ${\cal K}_2$ to fall off like $r^{-3}$ faster than $\Psi_0$ at large distances, ${\cal K}_2$ is of course not normalizable and such a solution generally exists.

\section{Discussion}

The observation that the leading asymptotic ground state wave function $\Psi_0$ is governed by supercharges for free superparticles has been pointed out previously in \cite{Polchinski:1999br, Hasler:2002wt}. This is perhaps obvious already from the perspective of effective field theory, though in the effective field theory approach it may not have been clear how to construct a systematic asymptotic expansion. In the well known perturbative computation of scattering at large impact parameters \cite{Becker:1997wh, Becker:1997xw, Becker:1997cp, Helling:1999js}, beyond one-loop order one encounters infrared divergences, which have been mostly ignored.\footnote{The point is that an IR divergence due to propagators at near zero frequency would have been cut off non-perturbatively, essentially due to the normalizability of the ground state wave function itself.}

The condition $P_0 Q^1_\A \Psi_0=0$ does not uniquely determine $\Psi_0$, however. If we had started with the wrong ansatz for $\Psi_0$, in principle there could be obstructions in solving the recursive equations for the asymptotic expansion at higher orders, or it could also be that the inconsistency is not visible at the level of the asymptotic expansion, but rather may be seen only after summing up the entire series in some way. It would also be tricky to guess a solution that is consistent with all symmetries of the problem. Our proposal is the simplest one that is consistent with all symmetries of the problem and the expected factorization property when the eigenvalues/D0-branes are divided into clusters on the Coulomb branch. There could be corrections to this proposal already at leading order, namely in $\Psi_0$ itself, but it does not seem easy to construct another solution with the desired symmetry properties. \cite{Hoppe:2000tj} suggested a different form of $\Psi_0$, which in principle could enter as a correction to our proposal, but it has a different scaling in $r$ and dies off faster at large distances. Even if such corrections are present in $\Psi_0$, it would not be possible to determine it based on the asymptotic expansion alone, as it would render $\Psi_0$ inhomogeneous under the overall scaling of $r$.\footnote{Note in particular that $r^{-14}$ is not an integer power of $r^{-{3\over 2}}$.}


The structure of the proposed $\Psi_0$ may provide some hints on the semi-classical nature of the bulk spacetime, at distances $r\ll N^{1\over 3}$ (in M-theory Planck units) from the origin. While $N^{1\over 7}\ll r\ll N^{1\over 3}$ is the weakly curved type IIA string theory regime, and $1\ll r\ll N^{1\over 7}$ is the weakly curved 11-dimensional M-theory regime, both lie in the strong 't Hooft coupling domain of the matrix quantum mechanics, and within the expected spatial spread of the ground state wave function. It has been mysterious why a probe eigenvalue that comes in from the asymptotic region (corresponding to a highly stringy regime in the bulk IIA picture) and interact with the ground state wave function of the remaining, say, $SU(N-1)$ part of the matrix quantum mechanics, would behave like a semi-classical particle governed by the Born-Infeld action in the bulk geometry. In our proposal for $\Psi_0$, which takes the form of a sum over products of two-body wave functions, one could hope the answer to be already approximately valid for $r_{ab}$'s that are parameterically large compared to 1 (or the scale set by $g_{YM}$ in the QM), as opposed to $N^{1\over 3}$ (or $N^{1\over 7}$ for that matter), though this is not at all obvious. Based on this form of $\Psi_0$ and its subleading corrections, perhaps a more reliable computation can be performed for the scattering of eigenvalues/D0-branes off the ground state wave function at impact parameters less than $N^{1\over 3}$, extending the results of \cite{Becker:1997xw, Helling:1999js} to the seemingly non-perturbative regime.\footnote{Even in the eikonal regime, taking into account the infrared modification of the propagators due to the ground state wave function already pollutes the structure of an analytic series expansion of the effective potential in $v^2/r^4$ and in $1/r^3$. This starts at $v^8$ order where the $r$ dependence is no longer fixed by supersymmetry.}


Eventually, we would like to count and understand the structure of long-lived metastable states of the matrix quantum mechanics at large $N$, which are supposed to be dual to microstates of the black hole in the bulk, either in the weakly coupled IIA regime or in the M-theory regime. Despite some numerical success based on Monte Carlo study of the thermal free energy \cite{Catterall:2007fp,Anagnostopoulos:2007fw, Catterall:2008yz,Hanada:2008gy,Catterall:2009xn,Hanada:2009ne}, there is little analytic understanding of the structure of such nonzero energy  states. Some encouraging results are obtained using truncated Schwinger-Dyson equations and extrapolating to the low temperature regime \cite{Kabat:1999hp, Kabat:2000zv, Kabat:2001ve, Lin:2013jra}. We hope a more precise understanding of the ground state wave function will provide insight on how to construct the general metastable excited states and ultimately a way to study Lorentzian observables relevant to the physics of black holes.


\bigskip

\section*{Acknowledgments}

We are grateful to Joe Polchinski for discussions. We would like to thank the organizers of the KITP program {\it New Methods in Nonperturbative Quantum Field Theory}, and the support of KITP during the course of this work. XY is supported by a Sloan Fellowship and by a Simons Investigator Award from the Simons Foundation.
This work is supported in part by the Fundamental Laws Initiative 
Fund at Harvard University, and by NSF Award PHY-0847457.

\appendix

\section{Change of variables in the asymptotic expansion}

Due to the constraint
\ie
\vec q_{ab}\cdot \vec r_{ab}=0,
\fe
we are only allowed to use
\ie
(\delta^{ij} - \widehat r^i_{ab} \widehat r^j_{ab}) {\partial\over \partial q^j_{ab}},
\fe
where $\widehat r_{ab} \equiv \vec r_{ab}/|\vec r_{ab}|$. 

Writing $X^i = U^{-1} (r^i_a E_a + q^i_{ab} T_{ab}) U$, we have
\ie\label{uxu}
U dX^i U^{-1} = [r^i_a E_a + q^i_{ab} T_{ab}, dU U^{-1}] + dr^i_a E_a + dq^i_{ab} T_{ab} .
\fe
Taking the trace of both sides multiplied by $(r^i_c-r^i_d)T_{dc} = r^i_{cd} T_{dc}$ (not summing over $c,d$), we have
\ie
r^i_{cd} \left( UdX^i U^{-1} \right)_{cd} &= r^i_{cd} {\rm Tr} \left( T_{dc}  [r^i_a E_a + q^i_{ab} T_{ab}, dU U^{-1}]  \right)
\\
&= |r_{cd}|^2 (dU U^{-1})_{cd} + r^i_{cd} \left[ q^i_{cb} (dU U^{-1})_{bd} - (dU U^{-1})_{cb} q^i_{bd} \right]
\fe
The second term on the RHS is down by a factor of $r^{-{3\over 2}}$ compared to the first term on the RHS, once we make the change of variables $q^i_{ab} = |r_{ab}|^{-{1 \over 2}} y^i_{ab}$ and maintain $y \sim \cO(1)$.
We can then express
\ie
& dU U^{-1} = \sum_{c\not=d} {r^i_{cd}\over r_{cd}^2} (U dX^i U^{-1})_{cd} T_{cd}
\\
&~~~~ - \sum_{c\not=d} {r^i_{cd}\over r_{cd}^2} \sum_{b\not=c,d} \left[{q^i_{cb} r^j_{bd}\over r_{bd}^2} (U dX^j U^{-1})_{bd} - {q^i_{bd} r^j_{cb}\over r_{cb}^2} (U dX^j U^{-1})_{cb} \right] T_{cd}
+{\cal O}(r^{-4})
\fe
Note that the diagonal components of $dU U^{-1}$ are unconstrained and are simply set to zero. Plugging this back into (\ref{uxu}), we have
\ie
& U dX^i U^{-1} 
= {r^i_{ab} r^j_{ab}\over r_{ab}^2} (U dX^i U^{-1})_{ab} T_{ab} + dr^i_a E_a + dq^i_{ab}T_{ab} 
+ {q^i_{ab} r^j_{ba} \over |r_{ab}|^2 } (U dX^j U^{-1})_{ba} (E_a-E_b)
\\
& \hspace{1in}
+ \sum_{c\not=a,b} \Pi^{ij}_{ab} \left[{q^j_{ac} r^k_{cb}\over r_{cb}^2} (U dX^k U^{-1})_{cb} - {q^j_{cb} r^k_{ac}\over r_{ac}^2} (U dX^k U^{-1})_{ac} \right] T_{ab} + {\cal O}(r^{-3}).
\fe

From this, we then solve for $dr^i_a$ and $dq^i_{ab}$ in terms of $dX^i$ up to ${\cal O}(r^{-3})$ terms.
\ie
& dr^i_a = {\rm Tr} (E_a U dX^i U^{-1}) + \sum_{b\not=a} {r^j_{ab}\over |r_{ab}|^{5\over 2} } \left[ y^i_{ab}  (U dX^j U^{-1})_{ba} + y^i_{ba} (U dX^j U^{-1})_{ab} \right] + {\cal O}(r^{-3}),
\\
& dq^i_{ab} = \Pi^{ij}_{ab}(U dX^i U^{-1})_{ab} \\
& \hspace{.5in} - \sum_{c\not=a,b} \Pi^{ij}_{ab} \left[{y^j_{ac}\over |r_{ac}|^{1\over 2}}{\widehat r^k_{cb}\over |r_{cb}|} (U dX^k U^{-1})_{cb} - {y^j_{cb}\over |r_{cb}|^{1\over 2}} {\widehat r^k_{ac}\over |r_{ac}|} (U dX^k U^{-1})_{ac} \right]  + {\cal O}(r^{-3}).
\fe


\bigskip
\begin{thebibliography}{}

\bibitem{Banks:1996vh} 
  T.~Banks, W.~Fischler, S.~H.~Shenker and L.~Susskind,
  ``M theory as a matrix model: A Conjecture,''
  Phys.\ Rev.\ D {\bf 55}, 5112 (1997)
  [hep-th/9610043].

\bibitem{Taylor:2001vb} 
  W.~Taylor,
  ``M(atrix) theory: Matrix quantum mechanics as a fundamental theory,''
  Rev.\ Mod.\ Phys.\  {\bf 73}, 419 (2001)
  [hep-th/0101126].

\bibitem{Maldacena:1997re} 
  J.~M.~Maldacena,
  ``The Large N limit of superconformal field theories and supergravity,''
  Adv.\ Theor.\ Math.\ Phys.\  {\bf 2}, 231 (1998)
  [hep-th/9711200].

\bibitem{Balasubramanian:1997kd} 
  V.~Balasubramanian, R.~Gopakumar and F.~Larsen,
  ``Gauge theory, geometry and the large N limit,''
  Nucl.\ Phys.\ B {\bf 526}, 415 (1998)
  [hep-th/9712077].

\bibitem{Susskind:1998vk} 
  L.~Susskind,
  ``Holography in the flat space limit,''
  hep-th/9901079.

\bibitem{Polchinski:1999br} 
  J.~Polchinski,
  ``M theory and the light cone,''
  Prog.\ Theor.\ Phys.\ Suppl.\  {\bf 134}, 158 (1999)
  [hep-th/9903165].

\bibitem{Becker:1997wh} 
  K.~Becker and M.~Becker,
  ``A Two loop test of M(atrix) theory,''
  Nucl.\ Phys.\ B {\bf 506}, 48 (1997)
  [hep-th/9705091].

\bibitem{Becker:1997xw} 
  K.~Becker, M.~Becker, J.~Polchinski and A.~A.~Tseytlin,
  ``Higher order graviton scattering in M(atrix) theory,''
  Phys.\ Rev.\ D {\bf 56}, 3174 (1997)
  [hep-th/9706072].

\bibitem{Becker:1997cp} 
  K.~Becker and M.~Becker,
  ``On graviton scattering amplitudes in M theory,''
  Phys.\ Rev.\ D {\bf 57}, 6464 (1998)
  [hep-th/9712238].


\bibitem{Helling:1999js} 
  R.~Helling, J.~Plefka, M.~Serone and A.~Waldron,
  ``Three graviton scattering in M theory,''
  Nucl.\ Phys.\ B {\bf 559}, 184 (1999)
  [hep-th/9905183].

\bibitem{Yi:1997eg} 
  P.~Yi,
  ``Witten index and threshold bound states of D-branes,''
  Nucl.\ Phys.\ B {\bf 505}, 307 (1997)
  [hep-th/9704098].

\bibitem{Sethi:1997pa} 
  S.~Sethi and M.~Stern,
  ``D-brane bound states redux,''
  Commun.\ Math.\ Phys.\  {\bf 194}, 675 (1998)
  [hep-th/9705046].

\bibitem{Moore:1998et} 
  G.~W.~Moore, N.~Nekrasov and S.~Shatashvili,
  ``D particle bound states and generalized instantons,''
  Commun.\ Math.\ Phys.\  {\bf 209}, 77 (2000)
  [hep-th/9803265].

\bibitem{Konechby:1998vc}
  A.~Konechny,
  ``On asymptotic Hamiltonian for SU(N) matrix theory,''
  JHEP\ {\bf 9810}, 18 (1998)
  [hep-th/9805046].

\bibitem{Porrati:1997ej}
  M.~Porrati, A.~Rozenberg,
  ``Bound states at threshold in supersymmetric quantum mechanics,''
  Nucl.\ Phys.\ {\bf B515}, 184-202 (1998)
  [hep-th/9708119]

\bibitem{Sethi:2000zf}
  S. Sethi and M.~Stern,
  ``Invariance theorems for supersymmetric Yang-Mills theories,''
  Adv.\ Theor.\ Math.\ Phys.\ {\bf 4}, 487-501 (2000)
  [hep-th/0001189]
  
\bibitem{Plefka:1997xq} 
  J.~Plefka and A.~Waldron,
  ``On the quantum mechanics of M(atrix) theory,''
  Nucl.\ Phys.\ B {\bf 512}, 460 (1998)
  [hep-th/9710104].
  
\bibitem{Frohlich:1999zf} 
  J.~Frohlich, G.~M.~Graf, D.~Hasler, J.~Hoppe and S.~-T.~Yau,
  ``Asymptotic form of zero energy wave functions in supersymmetric matrix models,''
  Nucl.\ Phys.\ B {\bf 567}, 231 (2000)
  [hep-th/9904182].

\bibitem{Hoppe:2000tj} 
  J.~Hoppe and J.~Plefka,
  ``The Asymptotic ground state of SU(3) matrix theory,''
  hep-th/0002107.
  
\bibitem{Hasler:2002wt} 
  D.~Hasler and J.~Hoppe,
  ``Asymptotic factorization of the ground state for SU(N) invariant supersymmetric matrix models,''
  hep-th/0206043.

\bibitem{Kabat:1999hp} 
  D.~N.~Kabat and G.~Lifschytz,
  ``Approximations for strongly coupled supersymmetric quantum mechanics,''
  Nucl.\ Phys.\ B {\bf 571}, 419 (2000)
  [hep-th/9910001].

\bibitem{Kabat:2000zv} 
  D.~N.~Kabat, G.~Lifschytz and D.~A.~Lowe,
  ``Black hole thermodynamics from calculations in strongly coupled gauge theory,''
  Int.\ J.\ Mod.\ Phys.\ A {\bf 16}, 856 (2001)
  [Phys.\ Rev.\ Lett.\  {\bf 86}, 1426 (2001)]
  [hep-th/0007051].
  
\bibitem{Kabat:2001ve} 
  D.~N.~Kabat, G.~Lifschytz and D.~A.~Lowe,
  ``Black hole entropy from nonperturbative gauge theory,''
  Phys.\ Rev.\ D {\bf 64}, 124015 (2001)
  [hep-th/0105171].

\bibitem{Lin:2013jra} 
  Y.~-H.~Lin, S.~-H.~Shao, Y.~Wang and X.~Yin,
  ``A Low Temperature Expansion for Matrix Quantum Mechanics,''
  arXiv:1304.1593 [hep-th].

\bibitem{Catterall:2007fp}
S. Catterall and T. Wiseman, 
``Towards lattice simulation of the gauge theory duals to black holes and hot strings,''
JHEP\ {\bf 0712} 104 (2007)
[arXiv:0706.3518 [hep-lat]].

\bibitem{Anagnostopoulos:2007fw} 
  K.~N.~Anagnostopoulos, M.~Hanada, J.~Nishimura and S.~Takeuchi,
  ``Monte Carlo studies of supersymmetric matrix quantum mechanics with sixteen supercharges at finite temperature,''
  Phys.\ Rev.\ Lett.\  {\bf 100}, 021601 (2008)
  [arXiv:0707.4454 [hep-th]].

\bibitem{Catterall:2008yz}
S. Catterall and T. Wiseman, 
``Black hole thermodynamics from simulations of lattice Yang-Mills theory,''
Phys.\ Rev.\ D\ {\bf 78} 041502 (2008)
[arXiv:0803.4273 [hep-th]].

\bibitem{Hanada:2008gy} 
  M.~Hanada, A.~Miwa, J.~Nishimura and S.~Takeuchi,
  ``Schwarzschild radius from Monte Carlo calculation of the Wilson loop in supersymmetric matrix quantum mechanics,''
  Phys.\ Rev.\ Lett.\  {\bf 102}, 181602 (2009)
  [arXiv:0811.2081 [hep-th]].

\bibitem{Catterall:2009xn}
S. Catterall and T. Wiseman, 
``Extracting black hole physics from the lattice,''
JHEP\ {\bf 1004} 077 (2007)
[arXiv:0909.4947 [hep-th]].

\bibitem{Hanada:2009ne} 
  M.~Hanada, J.~Nishimura, Y.~Sekino and T.~Yoneya,
  ``Monte Carlo studies of Matrix theory correlation functions,''
  Phys.\ Rev.\ Lett.\  {\bf 104}, 151601 (2010)
  [arXiv:0911.1623 [hep-th]].



\end{thebibliography}
\end{document}